\pdfoutput = 1
\RequirePackage{nag}
\RequirePackage{amssymb}
\RequirePackage{amssymb}
\documentclass[runningheads]{raa}
\usepackage[pagebackref=true,bookmarks=false, colorlinks = true, linkcolor = green, anchorcolor = red, citecolor = blue, filecolor = red, urlcolor = red]{hyperref}
\usepackage{graphicx,times}
\usepackage{natbib}
\usepackage{multirow}
\RequirePackage{caption}[2012/03/25]

\bibpunct{(}{)}{;}{a}{}{,}

\begin{document}

   \title{Time-resolved CCD photometry and time-series analysis of RR Lyrae type RR Gem star}
   \volnopage{Vol.0 (20xx) No.0, 000--000}\par\vspace{\baselineskip}
   \setcounter{page}{1}
   \author{V. A. Gohil and S.P. Bhatnagar}
   \institute{Department of Physics, M. K. Bhavnagar University, Bhavnagar - 364002, Gujarat, India; {\it spb@mkbhavuni.edu.in}\newline\newline
   \vs\no{\small Received~~2019 March 24; accepted~~2019~~April 09}}
   \date{}
   \abstract{ We present the results of a time-resolved photometric and time-series analysis of a RR Lyrae type star RR Gem. The main results are as follows: We found RR Gem's pulsation period, 0.39689 d, and its V and I mean magnitudes, 11.277 (V) and 11.063 (I) mag respectively. We confirm its variability type as RRab/BL because of its showing the Bla$\breve{z}$ko effect and it also shows asymmetric light curves (steep ascending branches), periods from 0.3 to 1.0 days, and amplitudes from 0.3 to 2 mag. in V. They are fundamental mode pulsators.
   }

   \maketitle

\section{Introduction}
\label{sect:intro}

RR Lyrae variables are older pulsating white giants with low metallicity. They are common in globular clusters-dense groups of old stars in the halos of galaxies. Like Cepheids, their pulsations are periodic. RR Lyraes have ~0.5 solar mass and have a short pulsation period of 0.2 to 0.85 days and amplitude variations of 0.3 to 2 magnitudes. RR Lyrae stars are usually spectral class A. The main sequence and the HB(Horizontal Branch) do not intersect. RR Lyrae stars are on the HB, indeed where the instability strip intersects the HB. The HB stars have left the red giant branch and are characterized by helium fusion in their cores surrounded by a shell of hydrogen fusion. All of the RR Lyrae stars in a cluster have the same average apparent magnitude. In different clusters, the average apparent magnitude is different. This is because all RR Lyraes have about the same average absolute magnitude of +0.75.\par\vspace{\baselineskip}

\noindent RR\footnote{\texttt{https://www.aavso.org/vsx/index.php?view=help.vartype\&nolayout=1\&abbrev=RR}} means that this is the RR variable in the Lyrae constellation, which are radially-pulsating giant A-F stars having amplitudes from 0.2 to 2 mag. in V band. Cases of variable light-curve shapes as well as variable periods are known. If their amplitudes are modified and also their periods, then they are called the "Bla$\breve{z}$ko effect"(\citealt{Blazko+1907})(denoted by the sub type BL). The majority of these stars belong to the spherical component of the Galaxy; they are present, sometimes in large numbers, in some globular clusters, where they are known as pulsating horizontal-branch stars. Like Cepheids, maximum expansion velocities of surface layers for these stars practically coincide with maximum light.

\newpage
\noindent RRAB/BL\footnote{\texttt{https://www.aavso.org/vsx/index.php?view=help.vartype\&nolayout=1\&abbrev=RRAB\%2FBL}} type variable stars are RR Lyrae variables with asymmetric light curves (steep ascending branches), periods from 0.3 to 1.0 days, and amplitudes from 0.5 to 2 mag. in V band. They are fundamental mode pulsators. BL indicates RR Lyrae stars showing the Bla$\breve{z}$ko effect.\par\vspace{\baselineskip}

\noindent J. Jurcsik and her team (\citealt{Jurcsik+Kovacs+1990, Jurcsik+Johanna+1998, Jurcsik+etal+2005, Jurcsik+etal+2006a, Jurcsik+etal+2006b, Jurcsik+etal+2006c, Jurcsik+etal+2008a, Jurcsik+etal+2008b, Jurcsik+2009a, Jurcsik+etal+2009b, Jurcsik+etal+2009c}) has done a remarkable work on the Blazhko behavior of RR Gem Star and in particular RR Lyr type stars showing Bla$\breve{z}$ko effects. Most RR Lyrae stars repeat their light curves with remarkable regularity. About 30 \% of the known galactic RR Lyrae stars, however, display cyclic modulation in shape and amplitude of the light curve over tens of pulsation cycles.

\section{Methodology}
\label{sect:meth}

\subsection{Instrumentation}

M. K. Bhavnagar University's Kumari Aanya Binoy Gaardi Observatory\footnote{\texttt{https://www.google.com/maps/place/Bhavnagar+Observatory/@21.7537555,72.1302003,\newline18z/data=!4m5!3m4!1s0x395f509a719b3a2b:0x30504b9860ad6057!8m2!3d21.7542085!\newline4d72.1303511?hl=en}} was used for this study. The Fully automated Observatory contains a 14 inch Celestron Schmidt-Cassegrain reflector type Telescope mounted on a German Equatorial mount. Telescope has an SBIG STF-7 CCD Camera along with SBIG CFW-8 filter wheel with Johnson-Cousins \emph{UBVRI}\footnote{\texttt{http://www.aip.de/en/research/facilities/stella/instruments/data/johnson\newline-ubvri-filter-curves}} photometric filters inbuilt into the CCD are attached on the telescope. Telescope is manual as well as fully computer controlled and telescope has two stepper motors for RA and DEC axis which is controlled by stepper motor controllers indigenously developed in collaboration with IUCAA, Pune(Inter University Center for Astronomy and Astrophysics). Open source software SCOPE is used to control the stepper motor controller and Cartes du Ciel(Sky Chart) was used for the CCD control as well as to give positions commands to SCOPE software which send command to stepper motor controller. Stepper motor controller performs the PWM and send electrical signals to RA and DEC stepper motors. There is an USB connection between CCD and PC for data as well as commands. Astronomy data is locally stored in Windows XP based PC. A GPS receiver is also connected with Windows PC which periodically updates time of the PC which in turn updates the DOC based PC's clock and thus the whole system's time gets updated  periodically. Manual sliding rooftop is connected with a motor through pulley for opening and closing down the Observatory's roof.

\subsection{Observations and data reduction}

\subsubsection{Differential Photometry}

\noindent MaxIm DL\footnote{\texttt{http://diffractionlimited.com/product/maxim-dl/}} photometric image analysis software was used for our data reduction work. Each and every time-series astronomical image was first calibrated for basic noise removal and then every continuous sequence was aligned and then photometry was performed on them.\par\vspace{\baselineskip}

\noindent For doing differential photometry(\citealt{Rodrigez+2005}) we needed our object star(RR Gem)\footnote{\texttt{https://www.aavso.org/vsx/index.php?view=detail.top\&oid=14304}}, one or more check stars and one reference star(standard star). Standard observatory procedures were used and used with minimum air-mass considerations.\par\vspace{\baselineskip}

\noindent For data reduction(calibration), flats field frames, dark frames and bias frames as well as light frames(See Figure 1.) were required so we programmed it into our MaxIm DL software and then it's done automatically according to our exposure time and other Johnson-Cousins ubvri filter value settings. For flat fielding we used daily twilight images and after all these we made master files of each of these and used them in our final calibration part.\par\vspace{\baselineskip}

\noindent In MaxIm DL we combined all the flats, darks and  bias files and combined them into single standard master files according to different photometry filers. For calibration we used these master files. We first opened all the object star's light images and loaded the master calibration files and performed calibrate using calibrate all menu. By this we get all the files calibrated, then we again opened the calibrated light images and align them using the align tool.\par\vspace{\baselineskip}

\noindent Now for performing Photometry we again opened all the light images and using the photometry menu's differential photometry tool we performed the photometry on the light images by selecting each individual object star, reference star and check stars. Finally it  gave us results in a Magnitude Vs JD(Jullian Day) graph in CSV or AAVSO Text format file.

\subsubsection{Time-series analysis}

\noindent Vstar(AAVSO)\footnote{\texttt{https://www.aavso.org/vstar}} program was used for performing time-series analysis (\citealt{Rodrigez+2005}). We used DC DFT with Period Range tool and fed the proper values in it like low period as 0.1 d, high period as 0.5 d and resolution as 0.0001 and we got the periodogram with period 0.39689 d.\par\vspace{\baselineskip}

\noindent We downloaded the standard published data for our star from AAVSO data tool\footnote{\texttt{https://www.aavso.org/data-download}} for the period of 10 months to get the bigger view of our star and performed the same above mentioned process and got the results(See Figures 2 and 3) then we also downloaded about 1 months data near to our observing period and analyzed the same and got this results(See Figures 4 and 5). By comparing this one month's data with our data(See Figures 6, 7 for (V) and 8, 9 for (I) respectively) we found it to be nearly perfectly matching with the magnitude as well as with the period. This ascertains that our observations and data reduction process is standardized.\par\vspace{\baselineskip}

%\newpage

\begin{figure}

\centering{\includegraphics[width=10cm]{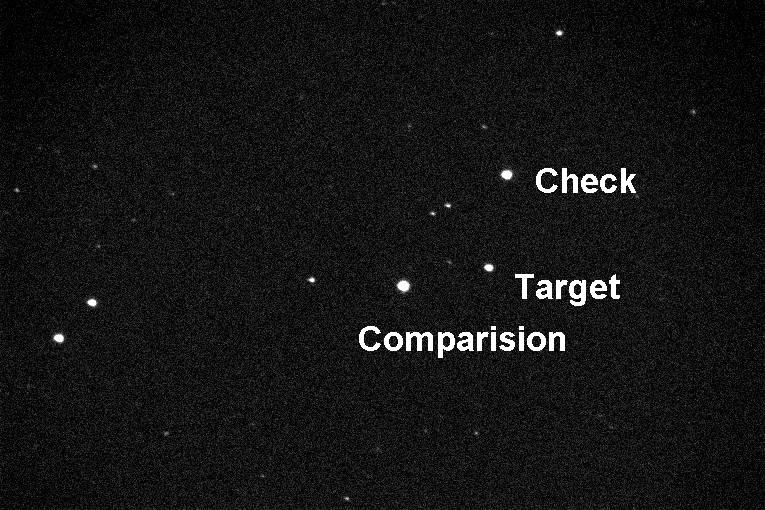}}

\caption{Clear Filter Object, Comparison and Check Star Field-CLEAR Filter: 16-04-2014\label{Fig1}}

\end{figure}

\begin{figure}
\centering{\includegraphics[width=10cm]{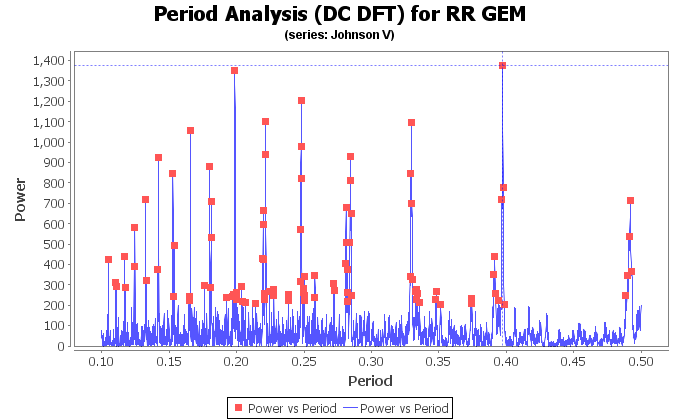}}

\caption{RR GEM 10 Months Data 2014 Johnson V Periodogram\label{Fig2}}

\end{figure}

\begin{figure}

\centering{\includegraphics[width=10cm]{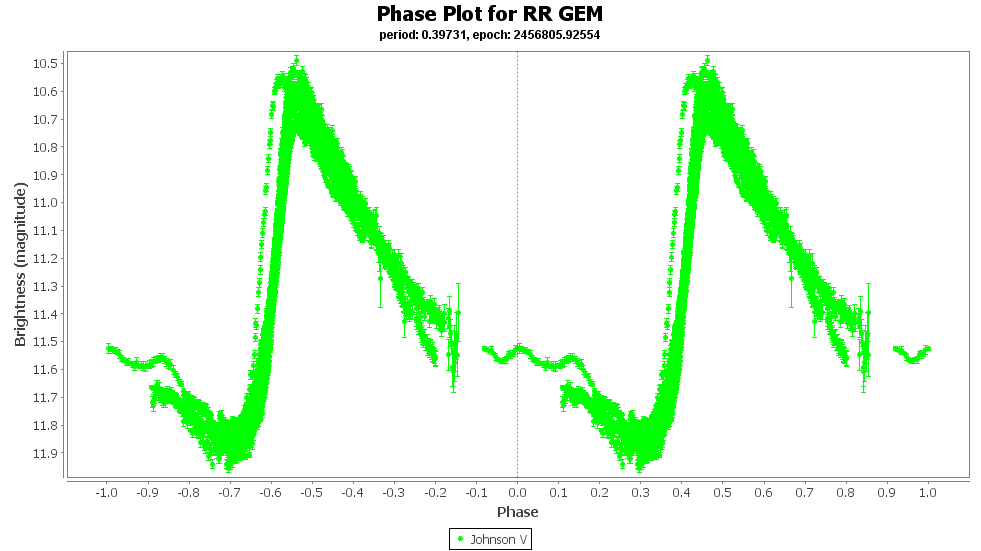}}

\caption{RR GEM 10 Months Phase Data 2014 Johnson V Phase Graph\label{Fig3}}

\end{figure}

\begin{figure}

\centering{\includegraphics[width=10cm]{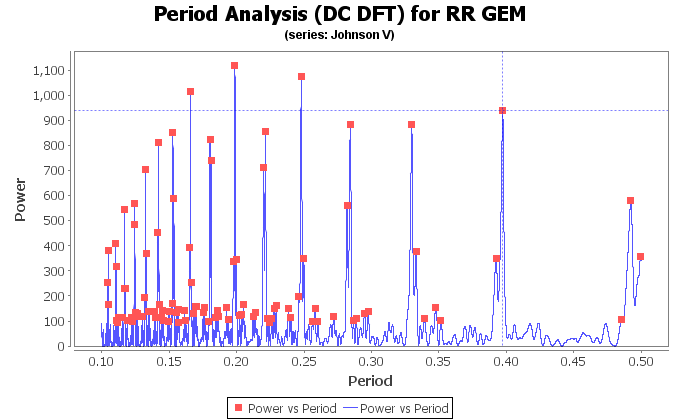}}

\caption{RR GEM 1 Months Data 2014 Johnson V Periodogram\label{Fig4}}

\end{figure}

\begin{figure}

\centering{\includegraphics[width=10cm]{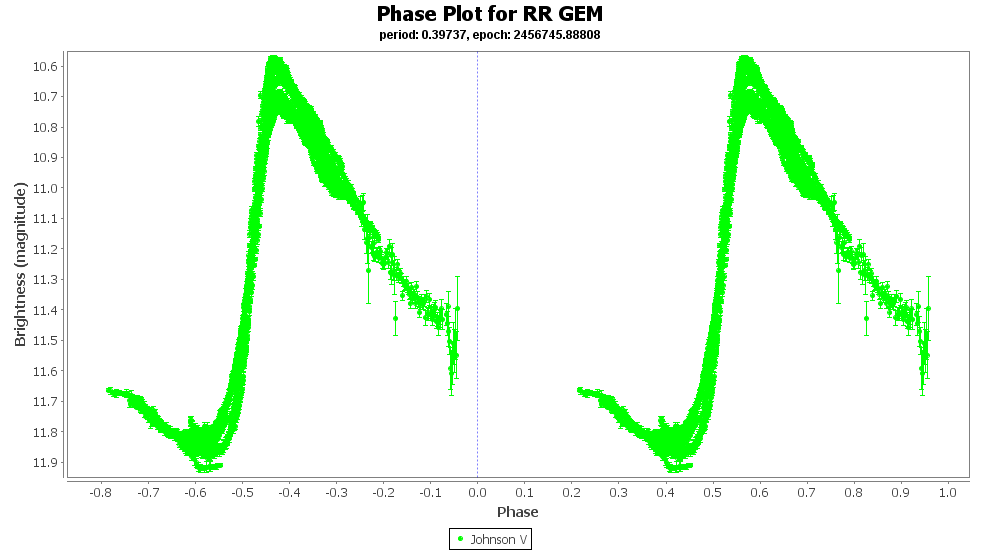}}

\caption{RR GEM 1 Months Phase Data 2014 Johnson V Phase Graph\label{Fig5}}

\end{figure}

\begin{figure}

\centering{\includegraphics[width=10cm]{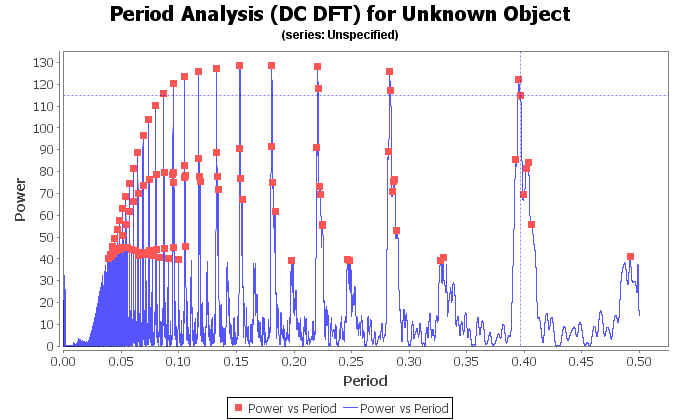}}

\caption{RR GEM 15 Days Our Data 2014 Johnson V Periodogram\label{Fig6}}

\end{figure}

\begin{figure}

\centering{\includegraphics[width=10cm]{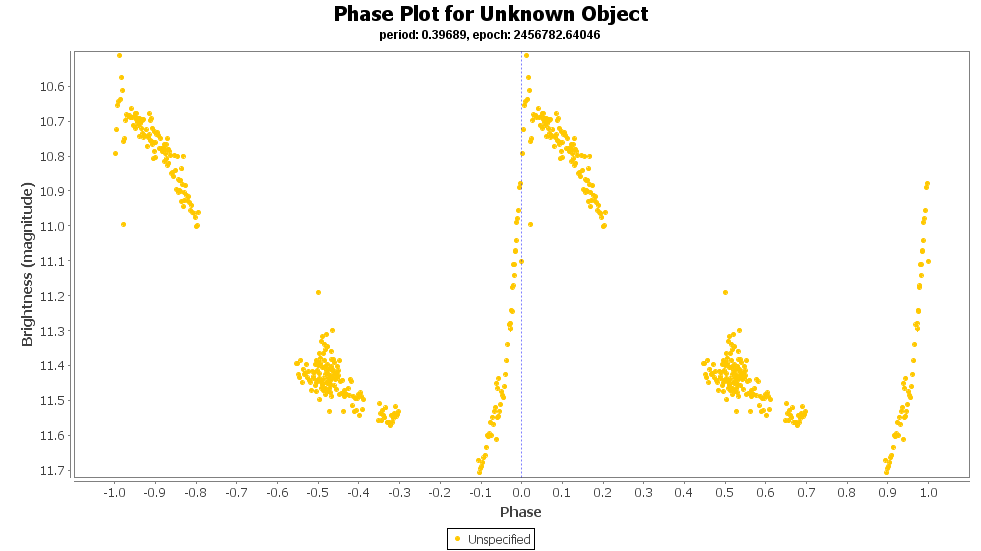}}

\caption{RR GEM 15 Days Phase Our Data 2014 Johnson V Phase Graph\label{Fig7}}

\end{figure}

\begin{figure}

\centering{\includegraphics[width=10cm]{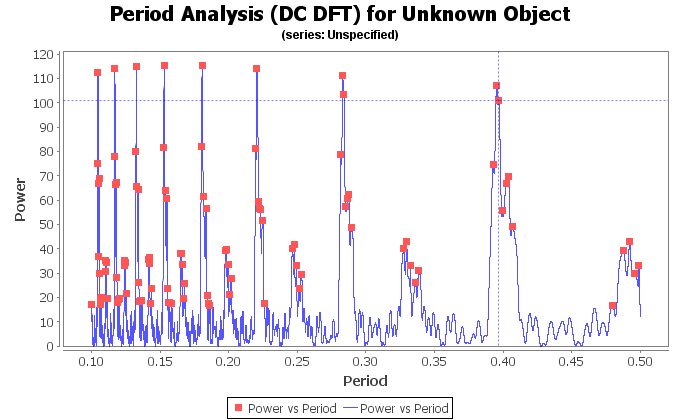}}

\caption{RR GEM 15 Days Our Data 2014 Johnson I Periodogram\label{Fig8}}

\end{figure}

\begin{figure}

\centering{\includegraphics[width=10cm]{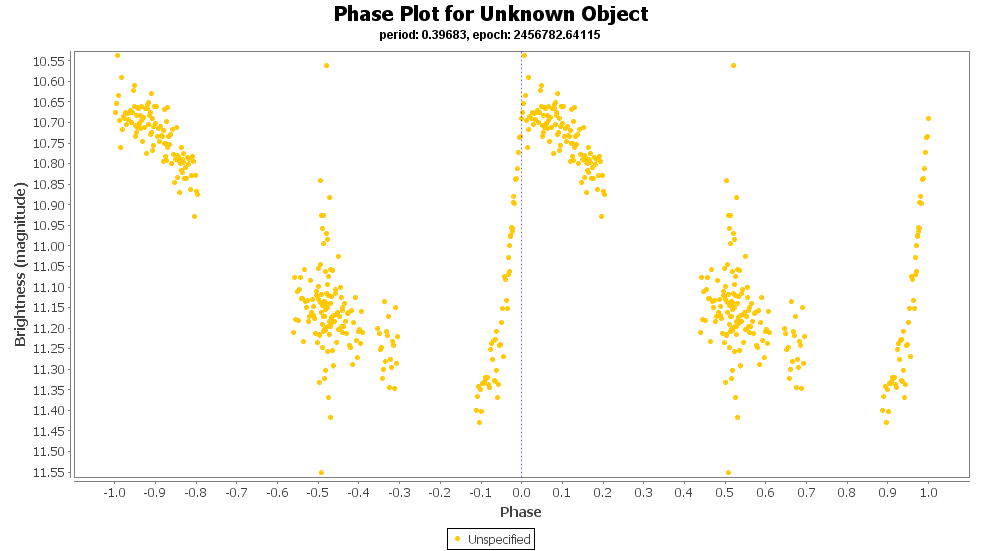}}

\caption{RR GEM 15 Days Phase Our Data 2014 Johnson I Phase Graph\label{Fig9}}

\end{figure}

%\clearpage
\newpage

\noindent According to AAVSO(VSX)\footnote{\texttt{https://www.aavso.org/vsx/index.php?view=search.top}} its spectral type is A9-F6. It was discovered by L. Ceraski (Ceraski, L. 1903) and all these information is based on 09 Feb 1972 (HJD 2441357.205). Its AAVSO UID : 000-BBM-940. Its other names are AAVSO 0715+31, AN 13.1903, HIP 35667. Our data is from JD: 2456764.16483 (16 APR 2014) to JD: 2456801.11748 (23 MAY 2014) totaling 15 nights. During this time the mean magnitudes in visual filter and infrared filters were 11.277 (V) and 11.063 (I) respectively.\par\vspace{\baselineskip}

\noindent We also compared our results with other astronomical databases. The details are as shown in Table 1.

\begin{table}[ht]
\centering
\resizebox{\textwidth}{!}{
\begin{minipage}{24.5cm}
\begin{tabular}{|c|c|c|c|c|c|c|c|c|c|c|c|c|}
\hline

\textbf{Database}& \textbf{ID / Name}&
\multicolumn{2}{|c|}{\textbf{J2000.0}}&
\textbf{Type}&   \textbf{V mag}&    \textbf{V mag}&  \textbf{Period}&  \textbf{Epoch}&  \textbf{Spectral Class}\footnote{\texttt{https://www.handprint.com/ASTRO/specclass.html}}\\
%\hline
&&($RA$)&($DEC$)&&($Max$)&($Min$)&($d$)&&\\
\hline

VSX\footnote{\texttt{Variable Star Index(AAVSO).}}&000-BBM-940&07:21:33.53&30:52:59.50&RRab/BL&10.62&11.99&0.3973106&2441357.205(HJD)&A9-F6\\
\hline
ASAS-SN\footnote{\texttt{The All Sky Automated Survey for SuperNovae.}}&ASASSN-VJ072133.66+305259.7&07:21:33.66&30:52:59.70&RRab&-&11.34&0.3992869&2457690.06(HJD)&-\\
\hline
CDS\footnote{\texttt{Strasbourg astronomical Data Center(CDS-Strasbourg).}}&V\^* RR Gem&07:21:33.53&30:52:59.46&Variable Star of RR Lyr type&-&11.92&-&-&A8\\
\hline
GCVS\footnote{\texttt{General Catalog of Variable Stars.}}
&RR Gem&07:21:33.53&30:52:59.50&RRab&10.62&11.99&0.3973106&41357.21&A9-F6\\
\hline
GVSSG\footnote{\texttt{General Variable Star Search Gateway.}}&RR Gem / 380010&07:21:33.50&30:52:59.50&RRab&10.62&11.99&0.3973106&41357.21(JD)&A9-F6\\
\hline
\end{tabular}
\end{minipage}}
\caption{RR Gem in different Astronomical databases.}
\end{table}

\noindent As shown in Table 1, we compared our results with many astronomical databases like VSX, ASAS-SN\footnote{\texttt{https://asas-sn.osu.edu/database/light\_curves/195828}},CDS\footnote{\texttt{http://cdsportal.u-strasbg.fr/?target=rr\%20gem}},GCVS\footnote{\texttt{http://www.sai.msu.su/gcvs/cgi-bin/search.cgi?search=RR+Gem}} and General Variable Star Search Gateway\footnote{\texttt{http://var2.astro.cz/gsg/index.php?star=rr+gem\&all=yes\&alldata=yes\&oejv=yes\&gcvs\newline=yes\&nsv=yes\&brka=yes\&meka=yes\&czev=yes\&bcvs=yes\&dssplate=yes\&usecoords=GCVS\&rezim=\newline search\_now}}.
\noindent We found our mean magnitude in V band: 11.277 mag and in I band: 11.063 mag, which is quite close to the published databases values like 11.185 mag in V band with the difference between our results and VSX(AAVSO) with the minor difference of 0.092, which is quite close.\par\vspace{\baselineskip}

\noindent We also compared our period with the astronomical databases and found that our period 0.39689 d is very close to the published value 0.3973106 d VSX(AAVSO) with the minor difference of 0.0004206 d, which is also quite close.
%\clearpage

\subsubsection{Comparision and Check Stars}\hfill

\noindent We used Check and Comparison Stars\footnote{\texttt{https://www.aavso.org/apps/vsp/photometry/?fov=18.5\&scale=F\&star=rr+gem\newline\&orientation=ccd\&maglimit=18.5\&resolution=150\&north=up\&east=left\&type=chart}} from Field photometry\footnote{\texttt{https://www.aavso.org/apps/vsp/photometry/?fov=18.5\&scale=F\&star=rr+gem\newline\&orientation=ccd\&maglimit=18.5\&resolution=150\&north=up\&east=left\&type=photometry}} for RR Gem from AAVSO Variable Star Database:

\begin{table}[ht]
\centering
\resizebox{\textwidth}{!}{
\begin{minipage}{21.5cm}
\begin{tabular}{|c|c|c|c|c|c|c|c|c|}

 \hline
 \textbf{Star type}& \textbf{AUID}&
 \multicolumn{2}{|c|}{\textbf{J2000.0}}&
 \multicolumn{5}{c|}{\textbf{Magnitudes}}\\

 &&($RA$)&($DEC$)&($B$)&($V$)&($B-V$)&($Rc$)&($Ic$)\\

 \hline
 Standard / Comparison Star&000-BJR-395&07:21:40.41&30:52:23.80&9.560(0.020)\^*&9.412(0.016)\^*&0.148(0.026)&9.322(0.020)\^*&9.219(0.021)\^*\\
 \hline
 Check Star&000-BBM-938&07:21:33.34&30:54:43&10.652(0.016)\^*&10.271(0.014)\^*&0.381(0.021)&10.043(0.017)\^*&9.826(0.019)\^*\\
 \hline
\end{tabular}
\end{minipage}}
\caption{Comparison and Check Star Table for RR Gem.}
\end{table}

\noindent The comparison / standard star and check stars were selected from the AAVSO's field photometry section. They were selected because they were very near, in the same CCD field.  They were chosen because they were about the same magnitude as well as were within the CCD field. All the details are given in the table 2. Practically they should be near to your object star and also their magnitude as well as spectral class should also match. Comparison star should be a standard non-variable star and check star could be non-variable or variable for the aperture(differential) photometry. So, they were chosen accordingly and we found them to be of satisfactory nature as per our requirements.

\section{Results}
\label{sect:res}

RR Gem's mean magnitude for V was 11.277 (V) mag and for I was 11.063 (I) mag. We had also compared star's data from AAVSO Variable star database for near about the same time period and found its visible filter's mean magnitude value to be 11.185 V, which gives the difference of 0.092 V which is also quite negligible.

\noindent After performing time-series analysis it was found that the period of RR Gem star published in the VSX(AAVSO) database was 0.3973106 d and after our analysis we found it to be 0.39689 d, the difference between the two is 0.0004206 d which is quite low. Thus it establishes our analysis matches with the published literature values.

\section{Conclusion}
\label{sect:concl}

We have established that our observations reported in this paper are in very good agreement with the published data of the star RR Gem and that it is an RR Lyrae type star and its variability type is RRab/BL and is clearly showing the Blazhko effect. We also conclude that our observatory can be now used for future studies of variable stars and in particular will be useful for performing optical light curve data and related analysis for the upcoming transient alerts like GW and Gamma-ray bursts which require optical followups from ground based telescopes during the bright phases (at various geographic longitudes).
%\clearpage

\begin{acknowledgements}
We would like to acknowledge the M. K. Bhavnagar University for providing Astronomical research facility at Kumari Aanya Binoy Gardi Observatory and We would also like to thank, IUCAA(Inter University Center For Astronomy And Astrophysics), AAVSO(American Association for Variable Star Observers), VSX(Variable Star Index)(AAVSO).
\end{acknowledgements}

\label{lastpage}


\begin{thebibliography}{99}

    \bibitem[Bla$\breve{z}$ko(1907)]{Blazko+1907} Bla$\breve{z}$ko, 1907, AN, 175, 325

    \bibitem[Jurcsik(1990)]{Jurcsik+Kovacs+1990} Jurcsik J., Kovacs G., 1990, A \& A, 312, 111

    \bibitem[Jurcsik(1998)]{Jurcsik+Johanna+1998} Jurcsik J., Johanna, 1998, A \& A, 333, 571

    \bibitem[Jurcsik et al.(2005)]{Jurcsik+etal+2005} Jurcsik J., S\'{o}dor, $\acute{A}$., V$\acute{a}$radi, M., Szeidl, B., Washuettl, A., Weber, M., D$\acute{e}$k$\acute{a}$ny, I., Hurta,\\ Zs., Lakatos, B., Posztob$\acute{a}$nyi, K., 2005, A \&A, 430, 1049

    \bibitem[Jurcsik et al.(2006a)]{Jurcsik+etal+2006a} Jurcsik J., Szeidl, B., S\'{o}dor, $\acute{A}$., D$\acute{e}$k$\acute{a}$ny, I., Hurta, Zs., Posztob$\acute{a}$nyi, K., Vida, K., V$\acute{a}$radi, M.,\\ Szing, A., 2006, AJ, 132, 61

    \bibitem[Jurcsik et al.(2006b)]{Jurcsik+etal+2006b} Jurcsik J., S\'{o}dor, $\acute{A}$., V$\acute{a}$radi, M., Vida, K., Posztobanyi, K., Szing, A., Hurta, Zs., Dekany, I.,\\ Washuettl, A., Vityi, N., 2006, \ibvs, 5709, 1

    \bibitem[Jurcsik et al.(2006c)]{Jurcsik+etal+2006c} Jurcsik J., Szeidl, B., V$\acute{a}$radi, M., Henden, A., Hurta, Zs., Lakatos, B., Posztob$\acute{a}$nyi, K.,\\ Klagyivik, P., S\'{o}dor, $\acute{A}$., 2006, A \& A, 445, 617

    \bibitem[Jurcsik et al.(2008a)]{Jurcsik+etal+2008a} Jurcsik J., S\'{o}dor, $\acute{A}$., Hurta, Zs., Kovari, Zs., Vida, K., Hajdu, G., Nagy, I., Dekany, I.,\\ Posztob$\acute{a}$nyi, K., Koponyas, B., 2008, \ibvs, 5844, 1

    \bibitem[Jurcsik et al.(2008b)]{Jurcsik+etal+2008b} Jurcsik J., S\'{o}dor, $\acute{A}$., Hurta, Zs., V$\acute{a}$radi, M., Szeidl, B., Smith, H. A., Henden, A., Dékány, I.,\\ Nagy, I., Posztob$\acute{a}$nyi, K., 2008, MNRAS, 391, 164

    \bibitem[Jurcsik(2009a)]{Jurcsik+2009a} Jurcsik J., 2009, CoAst, 159, 53

    \bibitem[Jurcsik et al.(2009b)]{Jurcsik+etal+2009b} Jurcsik J., S\'{o}dor, $\acute{A}$., Szeidl, B., Koll$\acute{a}$th, Z., Smith, H. A., Hurta, Zs., V$\acute{a}$radi, M., Henden, A.,\\ D$\acute{e}$kány, I., Nagy, I., 2009, MNRAS, 393, 1553

    \bibitem[Jurcsik et al.(2009c)]{Jurcsik+etal+2009c} Jurcsik J., S\'{o}dor, $\acute{A}$., Szeidl, B., Hurta, Zs., V$\acute{a}$radi, M., Posztob$\acute{a}$nyi, K., Vida, K., Hajdu, G.,\\ K\~{o}v$\acute{a}$ri, Zs., Nagy, I., 2009, MNRAS, 400, 1006

    \bibitem[Rodrigez(2005)]{Rodrigez+2005} P. Rodrigez-Gil, 2005, A \& A, 431, 289

\end{thebibliography}
\end{document}